 \documentclass[preprint2]{aastex}

\def\dsm{$M_\odot$}

\def\teff{$T_{\rm eff}$}

\shorttitle{The effects of rotation on the MSTO}
\shortauthors{Yang et al.}

\begin{document}


\title{THE EFFECTS OF ROTATION ON THE MAIN-SEQUENCE TURNOFF
OF INTERMEDIATE-AGE MASSIVE STAR CLUSTERS}

\author{Wuming Yang\altaffilmark{1,2}, Shaolan Bi\altaffilmark{1},
Xiangcun Meng\altaffilmark{2}, Zhie Liu\altaffilmark{1}}

\altaffiltext{1}{Department of Astronomy, Beijing Normal University,
Beijing 100875, China; yangwuming@bnu.edu.cn; \\ yangwuming@ynao.ac.cn}
\altaffiltext{2}{School of Physics and Chemistry, Henan Polytechnic
University, Jiaozuo 454000, Henan, China}

\begin{abstract}
The double or extended main-sequence turnoffs (MSTOs) in the
color-magnitude diagram (CMD) of intermediate-age massive star
clusters in the Large Magellanic Cloud are generally
interpreted as age spreads of a few hundred Myr.
However, such age spreads do not exist in younger
clusters (i.e., 40-300 Myr), which challenges this interpretation.
The effects of rotation on the MSTOs of star clusters have
been studied in previous works, but the results obtained are
conflicting. Compared with previous works, we consider the effects of
rotation on the MS lifetime of stars. Our calculations show
that rotating models have a fainter and redder MSTO with respect to
non-rotating counterparts with ages between about 0.8 and 2.2 Gyr,
but have a brighter and bluer MSTO when age is larger than 2.4 Gyr.
The spread of the MSTO caused by a typical rotation rate is equivalent to
the effect of an age spread of about 200 Myr. Rotation could lead to the double
or extended MSTOs in the CMD of the star clusters with
ages between about 0.8 and 2.2 Gyr. However, the extension
is not significant; and it does not even exist in younger clusters.
If the efficiency of the mixing were high enough,
the effects of the mixing would counteract the effect of the centrifugal
support in the late stage of \textbf{evolution}; and the rotationally
induced extension would disappear in the old intermediate-age
star clusters; but younger clusters would have an extended MSTO. \textbf{Moreover,}
the effects of rotation \textbf{might} aid in understanding the formation of
\textbf{some ``multiple populations''} in globular clusters.
\end{abstract}


\section{Introduction}
Since \cite{mack07} and \cite{mack08} discovered the double main-sequence
turnoffs (MSTOs) in the color-magnitude diagram (CMD) of intermediate-age
star clusters, such as NGC 1846, 1806, and 1783 in the Large Magellanic
Cloud (LMC), the phenomenon of the double or extended MSTOs has been
discovered in more and more star clusters \citep{mack08, glat08, milo09, gira09,
goud09, goud11a, kell12, piat13}. Some star clusters, such as NGC 1751 in the LMC
and NGC 419 \citep{milo09, gira09, rube11} in the Small Magellanic Cloud,
have not only the double or extended MSTOs, but also two distinct red clumps (RCs).
Recently, \cite{gira13} found that the NGC 411 in the Small Magellanic Cloud
also has an extended MSTO. Moreover, Omega Centauri exhibits a main-sequence (MS)
bifurcation \citep{piot05}; and NGC 2808 \citep{piot07} and NGC 6752 \citep{milo13b}
possess a triple MS split. These discoveries are not consistent with
the classical hypothesis that a star cluster is composed of stars belonging
to a simple, single stellar population with a uniform age and chemical composition.

The double or extended MSTOs have generally been interpreted as that the
population of star clusters have bimodal age distributions \citep{mack07, mack08}
or age spreads of about 100 - 500 Myr \citep{milo09, gira09, rube10, rube11,
kell12, piat13}. However, \cite{plat12} noted that this long period of star
formation seems to be odds with the fact that none of the younger clusters
are known to have such a trait. \cite{bast13} found that young massive
clusters NGC 1856 and 1866 in the LMC do not have such large age spreads,
which strongly queries the above interpretation of the age
spreads. The age is about 280 Myr for NGC 1856 \citep{kerb07} and
$\sim160-250$ Myr for NGC 1866 \citep{broc03}. Furthermore,
the age variation cannot be responsible for the MS spread in the young
cluster NGC 1844 with an age of about 150 Myr \citep{milo13a}.
Moreover, an age spread of about 460 Myr is required to explain
the dual RCs of NGC 1751 \citep{rube11}, which is two times longer than
$\sim$ 200 Myr for explaining the double MSTOs of NGC 1751 \citep{milo09}.
The contributions of interactive binary stars to the double
MSTOs and dual RCs were studied by \cite{yang11}. They found that binary
interactions and merging can reproduce the dual RCs and extended MSTO in the
CMD of an intermediate-age star cluster. However, the fraction of the interactive
binary systems is too low to explain the observed properties. In addition, the
binary interactions make the stars appear to have a younger age than non-interactive
systems; i.e. young stars are in the minority. This seems to be contrary
to the finding of \cite{milo09}, who found that about 70\% of stars belong to
the blue and bright (young) MSTO and around 30\% to the red and faint (old) MSTO.
The prediction of the interactive binaries seems to be contrary to this finding.

The effects of rotation on the structure and evolution of MS stars are
mainly of three types: (1) The effect of the centrifugal
support leads to a decrease in the effective temperature and luminosity
by decreasing the effective gravity.
(2) The mixing of elements can result in a decrease in the stellar radius
and an increase in the effective temperature by redistributing
chemical elements, compared to a non-rotating model at the same age.
However, the mixing can also lead to an increase in the
stellar radius and a decrease in the effective temperature by enlarging the
convective core of stars, compared to the non-rotating model at the same evolutionary
state (for example, at the end of the MS). (3) The Von Zeipel effect and the mixing
of elements can affect the instability of convection by changing the radiative
and adiabatic temperature gradients. If the convective core increases,
the lifetime of the MS will be prolonged, and vice versa
\citep{maed87, talo97, maed00, yang13a}.

The effects of rotation on the MSTO of intermediate-age star clusters
were first studied by \cite{bast09}, based on stellar models
computed with the code described by \cite{hege00} and \cite{brot11a}.
They found that rotating stars have a lower effective
temperature and luminosity than non-rotating counterparts and
concluded that stellar rotation can mimic the effect of a double population.
However, \cite{gira11} calculated the evolution of rotating models by using
the \cite{egge10} code and obtained that rotating models have a slightly
hotter and brighter turnoff with respect to non-rotating ones.
They concluded that the rotational effect could not explain
the presence of the extended MSTO. In the \cite{bast09} models,
the influence of rotation on the age of stars was neglected
and the effect of the rotationally induced mixing may be weaker
compared to that in \cite{gira11} models. This could lead
to the fact that their rotating models have a lower temperature
and luminosity than non-rotating ones in the whole stage of the MS.
Moreover, \cite{bast09} only calculated the evolution of a star with
the mass of 1.5 \dsm{}, assuming that the results for the star
can be applied to others. In fact, the effects of rotation on
the structure and evolution of stars depend on the mass of stars
for a given rotation rate \citep{maed00, yang13a}.
On the contrary, however, in the models of \cite{gira11},
the effects of the mixing may be more efficient,
which results in the fact that their rotating
models have a higher effective temperature than non-rotating ones
in the late stage of the MS for stars of any mass. In the early stage
of the MS of the \cite{egge10} models, the rotating models also exhibit
a lower effective temperature and luminosity, which mainly results
from the effect of the centrifugal support.
The latest evolutionary tracks calculated by \cite{geor13}
also show that the rotating models have a lower effective
temperature compared to the non-rotating ones in the early
stage of the MS (see their Fig. 11). If the extent of
the effects of mixing in the \cite{gira11} models
were different, the results might also be different.
Thus the effects of rotation on the MSTO need to be carefully
rechecked.

Furthermore, the observed characteristics of $\omega$ Centauri
and NGC 2808 can be interpreted by that a fraction of globular cluster stars
have sizable helium enhancements over the primordial value \citep{piot05,
piot07}. In addition, there are abundance anomalies in other globular
clusters \citep{grat04, milo13a}. In order to understand the self-enrichment
of globular cluster stars, \cite{vent01} and \cite{anto02}
suggested that the low-mass stars may have been polluted at the surface by accretion
from the gas that was lost from the evolving intermediate-mass asymptotic
giant branch stars, which requires a timescale of a few 100 Myr. However,
mass shedding models from fast rotating massive stars \citep{decr07} or
massive binaries \citep{mink09} can enrich the cluster on a timescale
of a few Myr. The rotational mixing in massive stars \citep{hunt08, brot11b}
and solar models \citep{yang07, bi11} is far from settled. Other
mixing processes may exist in the Sun and stars \citep{basu08,
hunt08, brot11b}, such as magnetic fields, gravity waves, and so on.

The evolutionary tracks of rotating models with
different rotation rates have been published by several
groups \citep{brot11a, brot11b, geor13}. However,
their calculations did not cover the mass range
between 1.3 and 1.7 \dsm{}. In this paper,
we also focus on the effects of rotation on the MSTO
of intermediate-age massive star clusters.
The paper is organized as follows: We simply show our stellar
evolutionary tracks in section 2; we present the results in section 3;
we discuss and summarize the results in section 4.

\section{Evolutionary tracks}
\subsection{Evolution code and assumptions}
We used the Yale Rotation Evolution Code \citep{pins89, yang07} to
compute the \textbf{evolution} of stellar models with and without rotation.
The input physics and some initial parameters are the same as
those used in \cite{yang07} and \cite{yang13a, yang13b}.
The meridional circulation, the Goldreich-Schubert-Fricke instability,
and the secular shear instability are considered \citep{enda78, pins89}
in all rotating models. The inhibiting effect of chemical gradients
on the efficiency of rotational mixing \citep{mest53} is also considered
in our models and is regulated by the parameter $f_{\mu}$
as described in \cite{pins89}. The angular momentum transport
and the mixing of elements caused by magnetic fields \citep{yang06}
and magnetic braking \citep{kawa88, chab95} are only
considered in stars with a mass less than 1.3 \dsm{}.
Angular momentum loss due to magnetic braking and mass loss
can be considered negligible in A-type stars \citep{macg94, wolf97, zore12}.
Thus in stars with mass larger than 1.3 \dsm{},
the angular momentum loss is not taken into account; i.e., the total
angular momentum is assumed to be conserved in these stars.
The transport of angular
momentum and the mixing of elements are treated as a diffusion
process in our models \citep{pins89, yang06}. The efficiency of
rotational mixing is described by the parameter $f_{c}$,
which is less than unity; and that parameter is used to
account for the fact that the instabilities mix material less
efficiently than they transport
angular momentum. Usually, the value of $f_{c}$ is about 0.02-0.03
\citep{chab92, yang06, hunt08, brot11b}. The value of
$f_{c}$ and $f_{\mu}$ is 0.0228 and 0.1 respectively
in the \cite{bast09} models (see \cite{brot11b}). In this work,
the values of $f_{c}$ and $f_{\mu}$ are both enhanced,
i.e., 0.2 for $f_{c}$ and 1.0 for $f_{\mu}$.

Centrifugal forces reduce the local effective gravity
at any point not on the axis of rotation, and they are generally
not parallel to the force of gravity. This leads to the facts
that equipotential surfaces of rotating stars
are no longer spheres and the radiative flux is not
constant on an equipotential surface. These effects
are taken directly into account in the equations of
hydrostatic equilibrium and radiative equilibrium
according to \cite{kipp70} and \cite{enda76}, as described
in \cite{enda76} and \cite{pins89}.

In our models, the initial metallicity $Z$ and hydrogen abundance
$X$ was fixed at 0.008 and 0.743, respectively. For a given mass,
the rotating and non-rotating models share the same initial parameters
except for the rotation rate. The rotating models were evolved
from the zero-age MS (ZAMS) to the end of the MS, assuming
the initially uniform rotation of $1.0\times10^{-4}$,
$1.5\times10^{-4}$, and $2.0\times10^{-4}$ radian s$^{-1}$ which
corresponds to an initial period ($P_{i}$) of about 0.73, 0.49, and 0.37 day
or an initial rotation rate ($\omega_{ini}$) of about 0.2, 0.3, and 0.4 times
the Keplerian rotation rate on ZAMS, respectively. The initial period of 0.49 day
produces surface velocities between about 130-180 km s$^{-1}$ for
stars with masses between 1.3-3.0 \dsm{} on ZAMS,
whose typical velocity is about 150 km s$^{-1}$ \citep{roye07}.
Compared with the initial velocity on the ZAMS of 150 km s$^{-1}$
in \cite{gira11} models, our initial velocity varies with mass.
In the simulation of \cite{bast09}, the mean of the initial rotation
rates is around 0.4 times the Keplerian rotation rate on the ZAMS.

\subsection{Evolutionary tracks}

\begin{figure}
\includegraphics[angle=-90, scale=0.34]{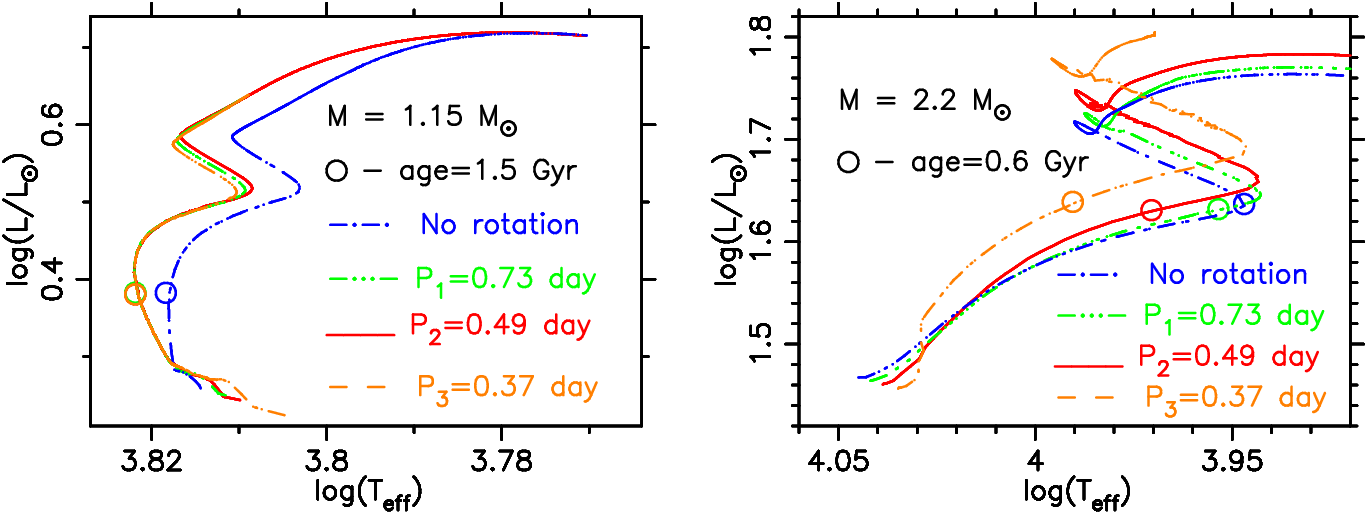}
\includegraphics[angle=-90, scale=0.34]{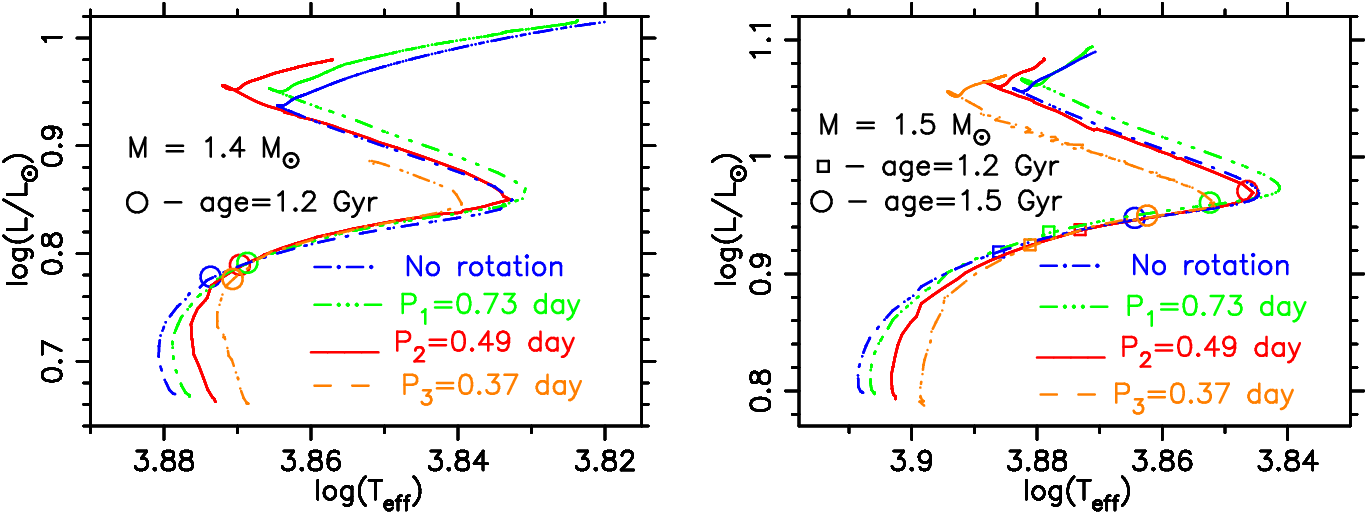}
\caption{Evolutionary tracks of rotating and non-rotating models in the
Hertzsprung-Russell (H-R) diagram. The value of $P_{i}(i = 1, 2, 3)$ is
the initial period on the ZAMS. The notation `$\circ$' shows a position of
the models with a given age. \label{fhr1}}
\end{figure}

\begin{figure}
\includegraphics[angle=-90, scale=0.5]{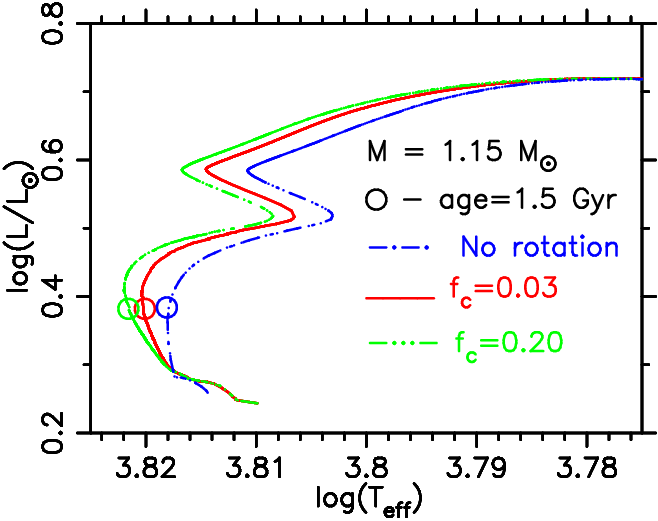}
\includegraphics[angle=-90, scale=0.5]{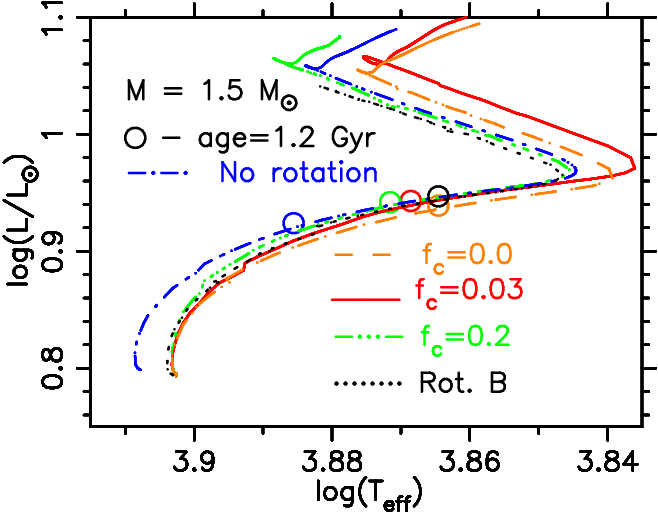}
\caption{Same as Fig. \ref{fhr1}. Rotating models
with the same initial period (0.49 day) but with different
efficiencies of element mixing ($f_{c}$). The evolution of
Rot B with $f_{c}=0.2$ was computed from the Wolff
ZAMS \citep{yang13b} to the end of the MS.
\label{fhrf}}
\end{figure}

The evolutionary tracks of rotating and non-rotating models
with $M=1.15, 1.4, 1.5$, and $2.2$ \dsm{} are shown
in Fig. \ref{fhr1}, where the positions of models with
a given age are labeled by the notation `o'.
For the star with $M=1.15$ \dsm{}, magnetic braking leads to
the effect that surface rotation velocity decreases
rapidly in a few hundred Myr. In this model, the
effect of the centrifugal support after the strong magnetic braking is not significant.
Thus the effects of rotation on this star are mainly dominated
by the rotational mixing. The rotating model mainly exhibits a higher effective
temperature than the non-rotating one at the same age.
The effects of rotation on the evolution of stars with mass less
than 1.3 \dsm{} are similar to those on the evolution of 1.15 \dsm{} star,
and they are more sensitive to the efficiency of the mixing than to
the rotation rate (see Figs. \ref{fhr1} and \ref{fhrf}).
For the star with $M=$ 2.2 \dsm{},
in the early stage of the MS phase the effect of centrifugal
support dominates the influences of rotation on the structure and
evolution of the star. Therefore, the rotating model has a lower
effective temperature and luminosity than the non-rotating one.
As the evolution proceeds, however, efficient mixing leads to the fact
that the outer edge of the convective core becomes slightly extended,
which enhances the effects of mixing \citep{maed00, yang13a}.
The efficient mixing prolongs the lifetime of the MS by feeding
fresh hydrogen fuel into the hydrogen-burning region
and makes the rotating models exhibit a higher effective temperature
than the non-rotating one at the same age (see Fig. \ref{fhr1}).
But at the end of the MS, the rotating
model exhibits a lower effective temperature and a higher
luminosity because the rotating models consumed more hydrogen
compared to the non-rotating ones. These are
consistent with the results of \cite{maed87} and \cite{egge10}.

For the stars with $M=$ 1.4 and 1.5 \dsm{},
in the early stage of the MS, due to the fast rotation and
the fact that the gradients of chemical compositions are small,
the effect of rotational mixing is not significant;
and the effect of the centrifugal support plays a dominant role.
Thus the rotating models have a lower luminosity and
effective temperature than their non-rotating counterparts.
But as the evolution proceeds, the rotating models can exhibit
a higher luminosity and a lower effective temperature than the
non-rotating one at the same age. The effective temperature
of the rotating models also can be higher than that of the non-rotating one,
depending on the age and rotation rate of the stars.

The effect of the centrifugal support leads
to a decrease in the effective temperature and gravity of stars.
The decrease of the gravity acts as an effective
decrease in the mass of the stars. The mass of the convective
core of MS stars decreases with decreasing stellar mass.
Therefore, the centrifugal effect results in a decrease in
the convective core. The
simulation of \cite{juli96} also shows that convective
penetration may be hindered in rotating stars.
Hydrogen is ignited at the same temperature.
The smaller the convective core, the less hydrogen
can be consumed in the core, and the shorter the lifetime of the MS.
Hence, the effect of the centrifugal support leads to an acceleration
of the evolution of stars. For example, the rotating
model with $M=$ 1.5 \dsm{} and $P=$ 0.49 day has evolved to
the MS hook at the age of 1.5 Gyr. However, the non-rotating
counterpart is still on MS with a higher effective temperature
and a lower luminosity (see Fig. \ref{fhr1}), which leads to the effect that
the rotating model seems older than the non-rotating one. In fact,
however, they have the same age.

The rotational mixing can bring hydrogen fuel into the core
from outer layers and transport the products of H-burning outwards.
The helium in the convective core is brought into the radiative region,
which leads to an increase in the
density $\rho$ at the bottom of the radiative envelope.
The adiabatic gradient $\triangledown_{ad}$ is proportional to $1/\rho$,
thus the convective core can be slightly increased by the mixing.
The increase brings more hydrogen fuel into the core and
enhances the effect of mixing. This prolongs the lifetime
of the MS of stars and leads to an increase in the effective
temperature by increasing the mean density of the stars as compared
to the non-rotating model at the same age.

The effects of rotation on the structure and evolution of stars
are a result of the competition between the effect of
the centrifugal support and that of rotational mixing. For stars
with mass larger than 1.3 \dsm{}, due to without magnetic braking
the effect of the centrifugal support plays a dominant role
in the early stage of the MS. In the late stage, however,
although the effect of the centrifugal support
plays an important role, the effects of mixing partly
counteract the influences of the centrifugal effect and become
even dominant. In the case of the evolution of rotating model
with $M=1.5$ \dsm{} and $P=0.37$ day, the effects
of mixing counteract the effect of the centrifugal support
in the late stage of the MS. Thus the luminosity and effective temperature
of the rotating model approximate those of the non-rotating one
at the age of 1.5 Gyr. But before the age of about 1.2 Gyr, the rotating
model has an obviously lower effective temperature compared to its
non-rotating counterpart. Our calculations show that
all stars with $Z=0.008$ and masses between around 1.3 and 2.0 \dsm{}
have similar characteristics.

We also calculated the evolutionary tracks of rotating models
with the initial rotation period of 0.49 day and a normal efficiency
of mixing ($f_{c}=0.03$ and $f_{\mu}=0.1$).
Figure \ref{fhrf} shows the evolutionary tracks of
models with $M=$ 1.15 and 1.5 \dsm{}. By way of comparison,
the track of the rotating model without mixing ($f_{c}=0$) of elements
but with other effects of rotation, such as the effect of the centrifugal
acceleration, is also plotted in the panel of the star with $M=$ 1.5 \dsm{}.
The lower the $f_{c}$, the smaller the effects of the
mixing. Thus the rotating models with a small $f_{c}$ have lower effective
temperatures compared to the rotating models with a large $f_{c}$.

\subsection{Comparison with Georgy's models}
\begin{figure}
\includegraphics[angle=-90, scale=0.5]{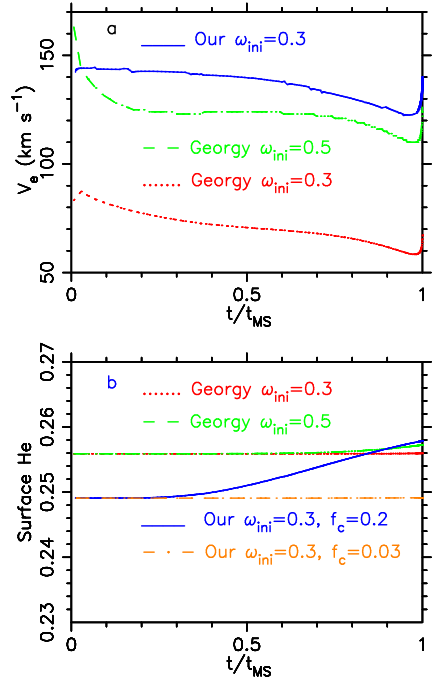}
\caption{Comparison between the surface velocity and He abundance
of our models with $M=1.7$ \dsm{} and those of \cite{geor13} with the same mass.
Panel $a$ shows the equatorial velocity as a function of the MS lifetime.
Panel $b$ indicates the evolution of the surface helium abundance.
The $\omega_{ini}$ shows the initial rotation rate.
\label{fher}}
\end{figure}

In panel $a$ of Fig. \ref{fher}, we compare the equatorial
velocity of our model with $M=1.7$ \dsm{} to the velocities of \cite{geor13}'s
models. The angular momentum loss is not taken into account in our models
with mass greater than 1.3 \dsm{}. But in the models of
\cite{geor13}, a substantial angular momentum was lost in the early
stage of the MS. Thus the equatorial velocity of our model
with $\omega_{ini}=0.3$ is obviously higher than that of \cite{geor13}'s
model with $\omega_{ini}=0.5$, with the result that
the effect of the centrifugal support appears to be
more significant in our model than in \cite{geor13}'s models with
the same initial rotation rate.

There is no obvious justification for adopting $f_{c}=0.2$.
The change in the mass fraction of the surface He caused by
rotational mixing is approximately equal to that of the surface hydrogen
in our models, which is consistent with \cite{geor13}'s calculation.
The efficient mixing could lead to the surface He enrichment.
Panel $b$ of Fig. \ref{fher} shows the \textbf{evolution} of surface He
abundance of our models and \textbf{that} of \cite{geor13}'s. The mass fraction
of surface He increases from the initial 0.249 to 0.258 at the end of
the MS in our rotating models with $f_{c}=0.2$ and $f_{\mu}=1$.
The value of 0.258 is compatible with the value of 0.257 in the Georgy's model.
The surface He abundance of rotating models with $f_{c}=0.03$
and $f_{\mu}=0.1$ is almost unchanged as the evolution proceeds
(see panel $b$ of Fig. \ref{fher}).

\section{Isochrones and synthetic results}
\subsection{Isochrones}
The rotating models of stars with masses between about 1.3 and
2.0 \dsm{} have a lower effective temperature than non-rotating ones,
which could lead to a spread in color of
intermediate-age massive star clusters. Thus we calculated a grid of
evolutionary tracks\footnote{The tracks can be obtained by e-mail
to Wuming Yang.} of rotating and non-rotating models with masses
between 1.0 and 3.0 \dsm{}. The mass interval $\delta M$ is
between 0.01 and 0.02 \dsm{} for the stars with masses between
1.15 and 2.0 \dsm{} but is about 0.1 \dsm{} for others.
The metallicity Z of evolutionary models was first converted
into [Fe/H]. Then the theoretical properties ([Fe/H], \teff{},
$\log g$, $\log L$) were transformed into colors and magnitudes
using the color transformation tables of \cite{leje98}.

\begin{figure}
\includegraphics[angle=-90, scale=0.34]{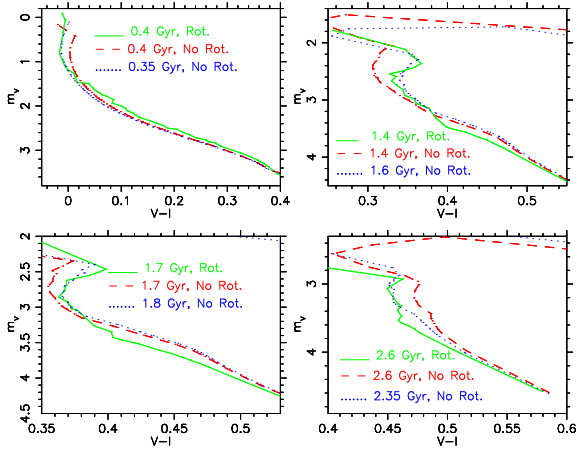}
\caption{The CMDs of the different isochrones. The solid (green) lines show
the isochrones of rotating models with an initial period of about 0.49 day
and $f_{c}=0.2$, while the dashed (red) and dotted (blue) lines indicate
the isochrones of non-rotating models with different ages.
 \label{fcmd}}
\end{figure}

Figure \ref{fcmd} shows the CMDs of different isochrones
obtained from our evolutionary models. All the rotating models
have the same initial rotation period of 0.49 day.
The isochrone shown by the dotted (blue) line has been
chosen by comparing the MSTOs of a set of isochrones with
an age interval of 50 Myr to those of the isochrones of rotating
models. The MSTO of rotating models with age = 1.4 Gyr
is nearly coincident with that of the non-rotating
models with age = 1.6 Gyr.
This indicates that the rotationally induced spread in color
is equivalent to the effect of an age spread of about
200 Myr. With the increase or decrease in age, the extension
becomes narrower and narrower. For example, when the age increases
to 1.7 or decreases to 0.9 Gyr, the extension is equivalent
to the effect of an age spread of about 100 Myr. This is because
the effect of the centrifugal acceleration is partly counteracted by
the effects of mixing. When the age is located between 0.8 and 1.9 Gyr,
rotating models exhibit a redder and fainter turnoff with respect to
their non-rotating counterparts, which leads to the effect
that rotating populations appear to be 100-200 Myr older
than non-rotating counterparts in the CMDs. In fact,
however, they have the same age. The \textbf{evolution}
of rotating models with masses between about 1.3 and 2.0 \dsm{}
\textbf{is} faster than \textbf{that} of non-rotating ones. Thus the mass of
the turnoff stars of the isochrone of rotating models
is slightly less than that of non-rotating ones.
Therefore, the MSTO of rotating
models is fainter than that of non-rotating ones.

When the age of a star cluster is older than 2.4 Gyr,
the rotating models exhibit a bluer and brighter
MSTO with respect to their non-rotating counterparts.
The rotationally induced spread is equivalent
to the effect of an age spread of about 200-400 Myr.
When the age of star clusters is younger than 0.6 Gyr,
the rotating models also exhibit a bluer and brighter MSTO
with respect to their non-rotating counterparts.
For example, in a star cluster with an age of 0.4 Gyr,
the spread caused by the rotation is similar to the effect
of an age spread of about 50 Myr. \cite{bast13} found
that in NGC 1856 and NGC 1866 the age spread is actually less than 35 Myr.
When the mass of stars is larger than about 2 \dsm{} or less
than about 1.3 \dsm{}, our rotating models mainly exhibit a higher
effective temperature than non-rotating ones at the same age.
Thus the rotating models have a bluer and brighter turnoff compared to
their non-rotating counterparts when age is younger than 0.6 Gyr or
older than 2.4 Gyr.

\begin{figure}
\includegraphics[angle=-90, scale=0.34]{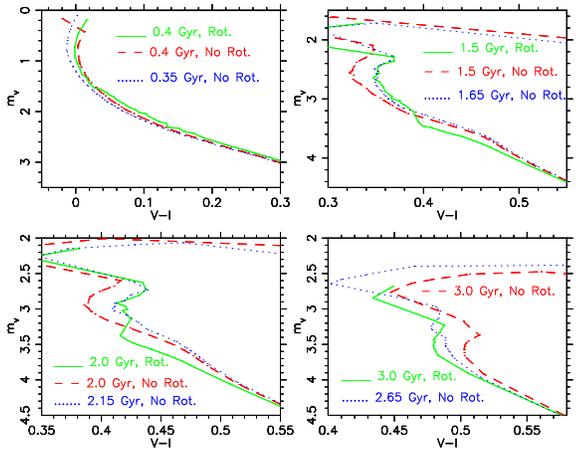}
\caption{Same as Fig. \ref{fcmd}, but the initial period of
rotating models here is about 0.73 day. The isochrone of
0.35 Gyr does not match the isochrone of rotating models, which is
plotted as a reference. \label{fcmdl}}
\end{figure}

The isochrones of rotating models with an initial period
of about 0.73 day are shown in Fig. \ref{fcmdl}. When the
age of star clusters is less than 0.7 Gyr,
the MSTO of isochrones of rotating models does not obviously
deviate from those of non-rotating counterparts.
When the age of star clusters is located between 0.9 and 2.2
Gyr, the rotating models have a redder and fainter MSTO compared
to the non-rotating counterparts. The spreads caused by the
rotation are equivalent to the effect of an age spread of about 100-200 Myr.
Compared to the extensions caused by the rotation with $P_{i}=0.49$ day,
these spreads appear in older intermediate-age star clusters;
but the results for the star clusters with age larger than 2.6 Gyr
are the same.

\begin{figure}
\includegraphics[angle=-90, scale=0.34]{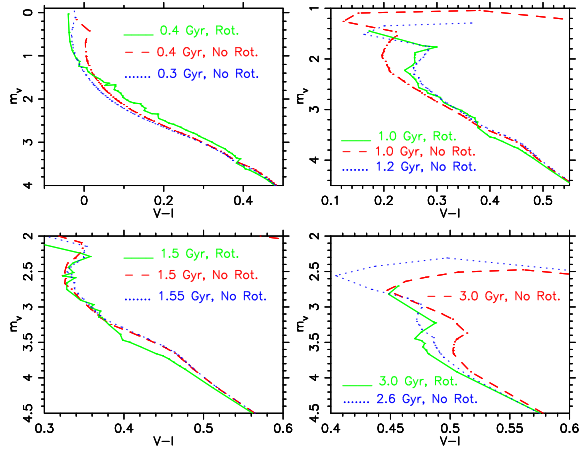}
\caption{Same as Fig. \ref{fcmd}, but the initial period of
rotating models here is about 0.37 day. \label{fcmdh}}
\end{figure}

Figure \ref{fcmdh} shows the results obtained
from the rotating models with an initial period of 0.37 day.
The interesting scenarios are that the spread caused by rotation
almost disappears in star clusters with ages between 1.4 and 2.0 Gyr,
but a spread similar to the effect of an age spread
of about 100-200 Myr still exists in the star clusters with ages
between 0.8 and 1.3 Gyr. When the age is less than 0.7 Gyr,
the rotating models have a bluer and brighter MSTO compared to
non-rotating ones, which is equivalent to the effect of an age
spread of about 100 Myr. This is because it is easier for
the effects of mixing to play a dominant role in the late stage of the MS
of stars for a high rotation rate (see Fig. \ref{fhr1}).
These results are similar to those of \cite{gira11}, except for star
clusters with ages between 0.9 and 1.2 Gyr. The MSTO of star clusters
with an age larger than 2.6 Gyr is similar to that of star clusters
with a lower rotation rate.

Comparing the isochrones obtained from rotating models \textbf{with $f_{c}=0.2$
but} with different rotation rates, one can find that the largest extension is
almost the same; i.e., it is approximately equivalent to the effect of an age
spread of about 200 Myr. For the intermediate-age star clusters, with
the increase in rotation rate the extension of the MSTO disappears in star
clusters with an older age \textbf{(i.e., about 1.5-2.0 Gyr)} due to the fact that
the effects of mixing counteract the effect of the centrifugal support.
For the star clusters with age larger than 2.4 Gyr,
the blue and bright extension of the MSTO is almost not
affected by the initial rotation rates because magnetic braking dominates
the rotation of stars with mass less than about 1.3 \dsm{}.
For young-age star clusters with age less than 0.6 Gyr,
an blue and bright extension of the MSTO can be produced by high rotation.

\begin{figure}
\includegraphics[angle=-90, scale=0.34]{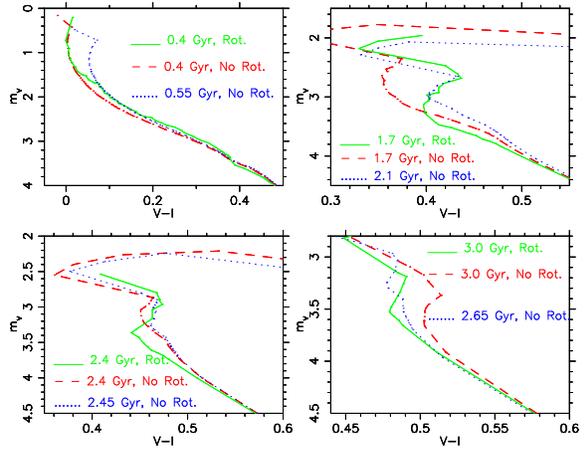}
\caption{Same as Fig. \ref{fcmd}, but the value of $f_{c}$ is 0.03.
The isochrone of 0.55 Gyr does not match the isochrone of rotating models,
which is plotted as a reference.
 \label{fcmdf}}
\end{figure}

The observed extension of the MSTO of intermediate-age
star clusters can be equivalent to the effect of an age spread of
about 100-500 Myr, which is broader than the extension caused by
rotating models with $f_{c}=0.2$. Figure \ref{fhrf} shows that
the effects of rotation on the evolution of stars are
sensitive to the value of $f_{c}$. Thus we calculated
the rotating \textbf{models with different rotation rates and
a lower efficiency of mixing ($f_{c}=0.03$ and $f_{\mu}=0.1$)}.
The results of this calculation \textbf{for $P_{i}=0.49$ day} are shown
in Fig. \ref{fcmdf}. When the age of star clusters is located
between 1.4 and 1.7 Gyr, the extension of MSTO caused by
rotation is equivalent to the effect of an age spread of
about 400 Myr. The extension decreases with increasing or
decreasing age. But the CMD of star clusters with age less than 0.6 Gyr is almost
not affected by rotation at all. When the age is larger than 2.6 Gyr,
rotating models exhibit a bluer and brighter MSTO compared to
non-rotating counterparts, which can be equivalent to the effect of
an age spread of about \textbf{350} Myr in the cluster with an age of 3.0 Gyr.

The extension of MSTO caused by rotation is mimicked by the
effect of an age spread of star clusters, as is shown in
Fig. \ref{fsp}. The positive spread shows that the isochrone of
rotating models has a redder MSTO, compared to that of non-rotating
models with the same age. The extension changes with the age of star
clusters, the initial rotation rate of stars, \textbf{and the value
of $f_{c}$;} and it is more significant in star clusters with age
between about 1-2 Gyr than in star clusters with age less than 0.6 Gyr.
\textbf{For $f_{c}=0.03$, due to the low efficiency of mixing,
the effect of mixing cannot compete with the effect of the centrifugal
support in stars with mass between about 1.3-2.0 \dsm{}. Thus the extension
caused by rotation with $f_{c}=0.03$ is more significant in star clusters 
with age between about 1-2 Gyr, compared with that caused by rotation 
with $f_{c}=0.2$. In addition, the extension increases with increasing
the initial rotation rate because the effect of the centrifugal support
increases with the increase in rotation rate (see the panels $b, d$,
and $f$ of Fig. \ref{fsp}).}

\subsection{Monte Carlo simulation}

In order to understand whether rotation can produce extended or
double MSTOs or not, we performed a stellar population synthesis
by way of Monte Carlo \textbf{simulations} for a total mass of about
$3\times10^{4}$ \dsm{} following the log-normal initial mass function
of \cite{chab01}. (The mass range of the intermediate-age massive star
clusters is between about (1-20)$\times10^{4}$ \dsm{} \citep{goud11b}.)
In order to match the results of \cite{milo09}, in which about
70\% of stars belong to the blue and bright MSTO and around 30\% to
the red and faint MSTO, we assumed that 70\% of stars
do not rotate, but that 30\% of stars rotate with the same initial
period of 0.49 day, though this assumption may not represent the
real situation. In our synthesized population,
we included observational errors taken to be a Gaussian distribution
with a standard deviation of 0.01 and 0.015 in magnitude
and color as in \cite{bast09}, respectively. The deviation of
0.015 in V-I corresponds to a deviation of about 90 K in effective
temperature in our models.

The CMDs of the synthesized star clusters
with different ages are shown in Fig. \ref{fvg}.
The double MSTOs can be clearly seen in the CMDs of the star clusters
with age = 1.2 and 1.5 Gyr, in which the hot and bright MSTO is dominant.
A continuously extended MSTO exists in the star cluster
with age = 0.9 Gyr, where the hot and bright MSTO is also dominant.
Compared with the extension of the MSTO of star clusters with
ages between 0.9 and 1.6 Gyr, that of the star clusters
with age = 0.4 and 1.7 Gyr is not obvious. Moreover, there is
almost no extension of the MSTO of the star cluster with age = 2.0 Gyr.
For the star clusters with age = 2.6 and 3.0 Gyr, their MSTO is
also extended by the effects of rotation, but the cool and faint
MSTO is dominant. The difference in the V-I between
the isochrone of rotating models and that of non-rotating
ones is less than 0.01 for the star cluster with the age of 0.4 Gyr.
Thus the extension of the MSTO in this cluster is dominated by the
deviation of 0.015 in color. However, the difference is about 0.04
for the star cluster with an age of 1.2 Gyr. The spreads in star
clusters of ages between 1.2 and 1.5 Gyr are mainly due to rotation.

The CMDs of the synthesized star clusters with different
rotation rates and $f_{c}$ are shown in Figs. \ref{fvgl}, \ref{fvgh},
and \ref{fvgf}, where the double or extended MSTOs can be
clearly seen in the CMD of star clusters with age between about 0.8-2.2 Gyr.

\textbf{The CMD of} NGC 1806 has double MSTOs, and its age is about
1400-1600 Myr \citep{milo09}. The initial rotation period of stars
in a star cluster should be different. For simplicity, we assumed that
the population of a star cluster is composed of 70\% non-rotating stars
and 30 \% rotating stars which is obtained by interpolating between
our rotating models with different rotation rates, assuming the
initial rotation rate has a Gaussian distribution with a peak at
about 0.3 times the Keplerian rotation rate on ZAMS and a standard deviation
of 0.06. However, this assumption may not represent the real distribution.
For example, the distributions of rotation rate of A0-A1 type stars
are \textbf{flatter} \citep{roye07}. A distance modulus of 18.45 and
$E(m_{\mathrm{F435}}-m_{\mathrm{F814}})$ of 0.14 are
adopted in this simulation; and the simulated results
are shown in Fig. \ref{fngc}, where the double MSTOs can be seen.
The simulated rotation population
reproduces well the extension of the MSTO of NGC 1806 observed by
\cite{milo09}. The age of the theoretical population is 1.4 Gyr,
and the value of $f_{c}$ is 0.2.

\section{Discussion and Summary}
\subsection{Discussion}
According to the studies of \cite{wolf04}, \cite{zore12}, and \cite{yang13b},
the ZAMS models of A-type stars should be a Wolff ZAMS model,
which is required for the \textbf{study} of the evolution of surface
velocity and the transport of angular momentum \citep{yang13b}.
For simplicity in computation, we adopted the ZAMS model with
uniform rotation. The rotating models computed in accord with the Wolff
ZAMS model also exhibit a lower effective temperature and higher
luminosity with respect to non-rotating ones in the late stage
of the MS of stars with masses between 1.3 and 2.0 \dsm{}. This is
consistent with the results obtained from the evolution of the
ZAMS model with a uniform rotation (see Fig. \ref{fhrf}). Thus
our results about the effects of rotation on the CMDs cannot be
significantly changed by the ZAMS model.

Our ZAMS models with mass larger than 2.0 \dsm{} have had a slight
gradient of interior elements as compared with lower mass stars,
in which the gradient can be completely neglected. Thus the mixing
of elements takes place at the beginning of evolution
for the stars with mass larger than 2.0 \dsm{}, but it acts in a later time
for stars with mass less than 2.0 \dsm{}. This leads to the conclusion that
the early evolution of the MS of stars with mass less than 2.0 \dsm{} is
readily dominated by the effect of centrifugal acceleration.
Moreover, for a given rotation rate,
the coefficient of the diffusion caused by rotation increases
with increasing mass; i.e. the efficiency of the mixing increases
with increasing mass. These lead to the result that
there is a critical mass $M_{c}$ for the effects of rotation
on the structure and evolution of the stars. For stars with a
mass larger than $M_{c}$, the effect of the mixing soon exceeds
the centrifugal effect in the early stage of the MS.
For stars with masses between about 1.3 \dsm{} and $M_{c}$, however,
the effect of the centrifugal acceleration is dominant for a long time
during the MS stage. In our models, the critical mass is
about 2.0 \dsm{} which is close to the critical mass of the He-flash
of \cite{yang12}, and it could be affected by
the mixing processes. Thus the value of about 2.0 \dsm{} is debatable.

In rotating stars, meridional circulation is an advection process.
The rotational mixing could be much more complex than that described
by a diffusion process. The value of 0.2 of $f_{c}$ in our models
is obviously higher than that calibrated against the nitrogen abundance
in massive stars (0.0228) \citep{hunt08, brot11b} or against the solar model
(0.03) \citep{yang06}. There may be other mixing mechanisms undescribed by the
present model in stars and sun \citep{hunt08, basu08}, such as gravity
waves, which are efficient at transporting angular momentum \citep{zahn97,
talo05}. The value of 0.2 for the $f_{c}$ corresponds to a high
efficient mixing in stars, which results in a higher mean density
for stars with mass less than 2.0 \dsm{} when they approach
the end of MS; i.e., these models with a high efficient mixing have
a lower radius. This helps explain the evolution of surface
velocity of the stars with mass less 2.0 \dsm{} \citep{zore12, yang13b}.
However, if the mixing is too efficient, this would introduce
an amount of hydrogen into the burning region of
the star, with a result that the extension of the MSTO caused
by rotation would disappear. Our current understanding of the mixing
processes could be limited. The mixing in stars is debatable.

The rotationally induced extension of MSTO does not exist
in the simulated clusters with age between 1.4 and 2.0 Gyr
when the initial rotation period is 0.37 day, which is similar
to the results of \cite{gira11}. In addition,
the extension also disappears in the young clusters with
age less than 0.7 Gyr and $P_{0}=0.73$ day. If we took a
fixed rotation velocity $V_{0}$ as the initial rotation
parameter for all models, the high mass stars would have
a lower rotation rate, but the low mass stars would have a
higher rotation rate. In that scenario, the result might
have been that rotation does not obviously affect the MSTO
of star clusters. Moreover, the efficiency of mixing would
be different between our models and \cite{gira11} models
because there is no adjustable parameter $f_{c}$ in
the Geneva stellar evolution code \citep{maed00}.
The different initial rotation parameter and efficiency of
mixing would lead to differences between our results and
\cite{gira11}'s. In addition, if \cite{gira11} considered
the angular momentum loss as \cite{geor13}
for stars with mass larger than 1.3 \dsm{}, the effect of mixing
would be easier to exhibit in their models than the effect of the
centrifugal support; i.e., their models would readily exhibit a higher
effective temperature compared with the non-rotating counterparts.

Compared, moreover, with \cite{bast09} models, we calculated a
dense grid of evolutionary tracks of rotating models and
considered the effects of rotation on the lifetime of the MS in
our isochrones and simulations. Our results are similar to
that of \cite{bast09}; i.e., the MSTO of star clusters with
ages between about 0.8 and 2.2 Gyr can be extended by
rotation. But in our models, the extension is
dependent on rotation rate,
the efficiency of mixing, and the age of star clusters.

In the 16 star clusters studied by \cite{milo09} (see
their Table 3), the star clusters with ages between 950 and 1400
Myr have an extended MSTO of about 150-250 Myr,
but the star clusters with age larger than 1550 Myr have
a smaller extension ($\Delta$age $< 100$ Myr). This seems to
be similar to our result that rotationally induced extension
decreases with the increase or decrease in the age of star
clusters. It seems, moreover, to be more similar to the
results obtained from high-efficiency models than to those obtained from
low-efficiency models.

The extension of the MSTO caused by rotation is significantly affected
by the efficiency of mixing and decreases with increasing the efficiency.
The star cluster SL 529, with an age of about 2.0 Gyr, has an \textbf{apparent}
age spread of about 500 Myr \citep{piat13}. The rotationally induced extension of
the MSTO of star cluster with an age of 2.0 Gyr is similar to the effect
of an age spread of about \textbf{350} Myr in our models with $f_{c}=0.03$
\textbf{and $P_{i}=0.37$ day. While the age of star clusters decreases to
1.7 Gyr, the extension is equivalent to the effect of an age spread of about
450 Myr (see the panel $f$ of Fig. \ref{fsp}).} The spread of star cluster
SL 529 seems not to be conflicting with our results.

\textbf{Stars with different rotational parameters and masses could
evolve to the stage of the core helium burning at the same time.
Thus the effects of rotation should contribute to the formation of
the dual red clumps. Our current rotating models cannot be computed
to the core-He-burning stage. Thus we do not obtain the effects
of rotation on the red-clump stars.}

\textbf{In addition, the MSTO of old-age star clusters also can be extended and
the surface He abundance of stars can be changed by the effects of rotation.
Thus the effects of rotation might have some contributions to the
formation of ``multiple populations" in globular clusters.}

\subsection{Summary}
For stars with masses between about 1.3 and 2.0 \dsm{},
the effect of the centrifugal support plays a dominant role
in the early stage of evolution, which leads to the fact
that rotating models have a lower effective temperature
and evolve faster than non-rotating ones. As the
\textbf{evolution proceeds}, however, the effect of rotational mixing
partly counteracts the influence of the centrifugal acceleration
and even dominates the effects of rotation on the structure
and evolution of stars. As a consequence, the MSTO of
intermediate-age star clusters with ages between 0.8 and
2.2 Gyr can extend redward, but the extension
decreases with increasing or decreasing age. In
rotating stars with mass less than 1.3 \dsm{}, the
effect of rotational mixing plays a dominant role during
the MS stage, which results in the fact that rotating models
have a higher effective temperature than non-rotating ones
at the same age. Thus the MSTO of the older star clusters
extends blueward. The \textbf{evolution} of rotating stars with
mass larger than 2.0 \dsm{} \textbf{is} mainly affected by
the effect of rotational mixing which depends on rotation
rate and the efficiency of mixing. Therefore, the MSTO of
young star clusters can extend slightly blueward or
cannot extend at all, dependent, likewise,
on the rotation rate and efficiency of the mixing.

\begin{figure}
\includegraphics[angle=-90, scale=0.34]{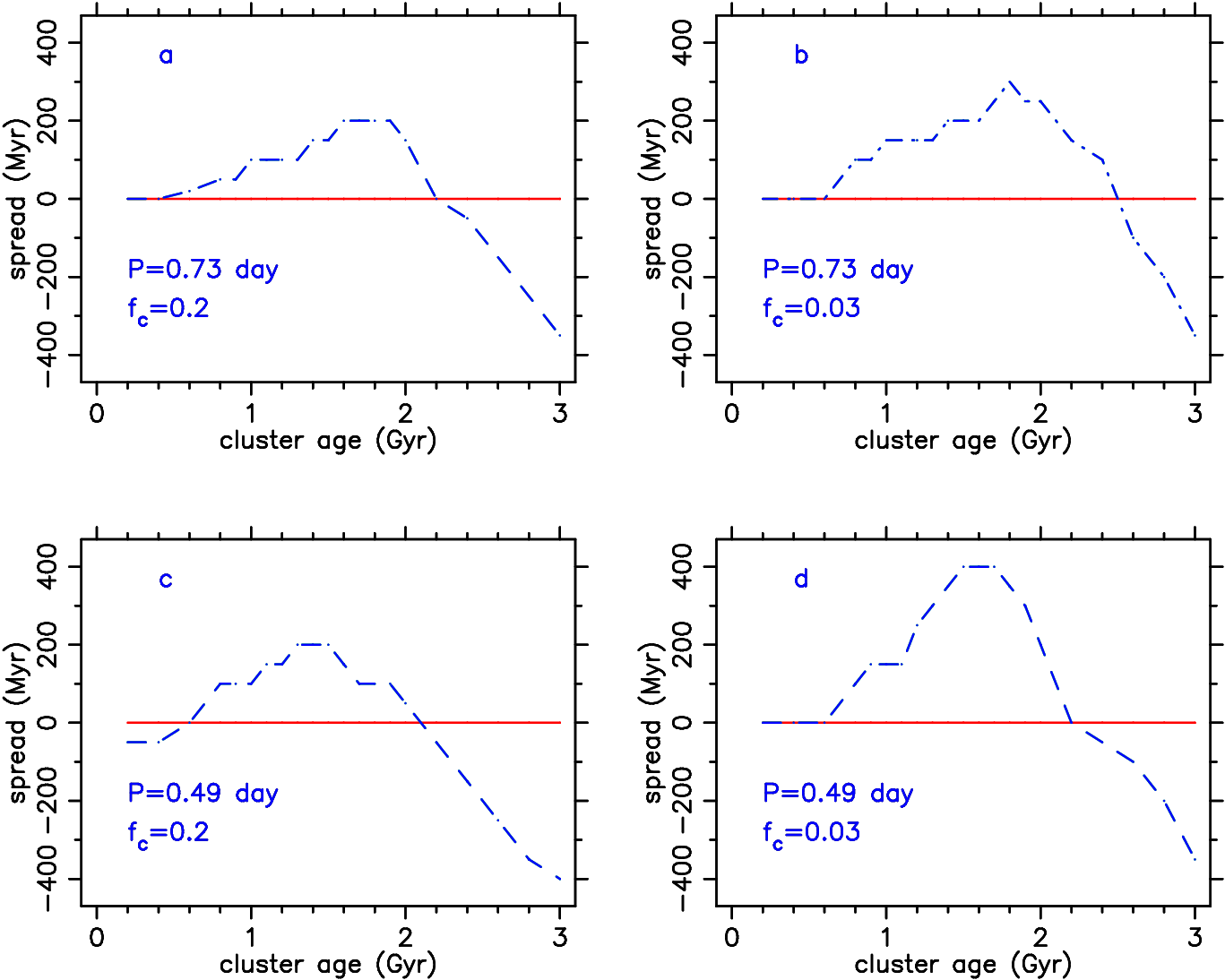}
\includegraphics[angle=-90, scale=0.34]{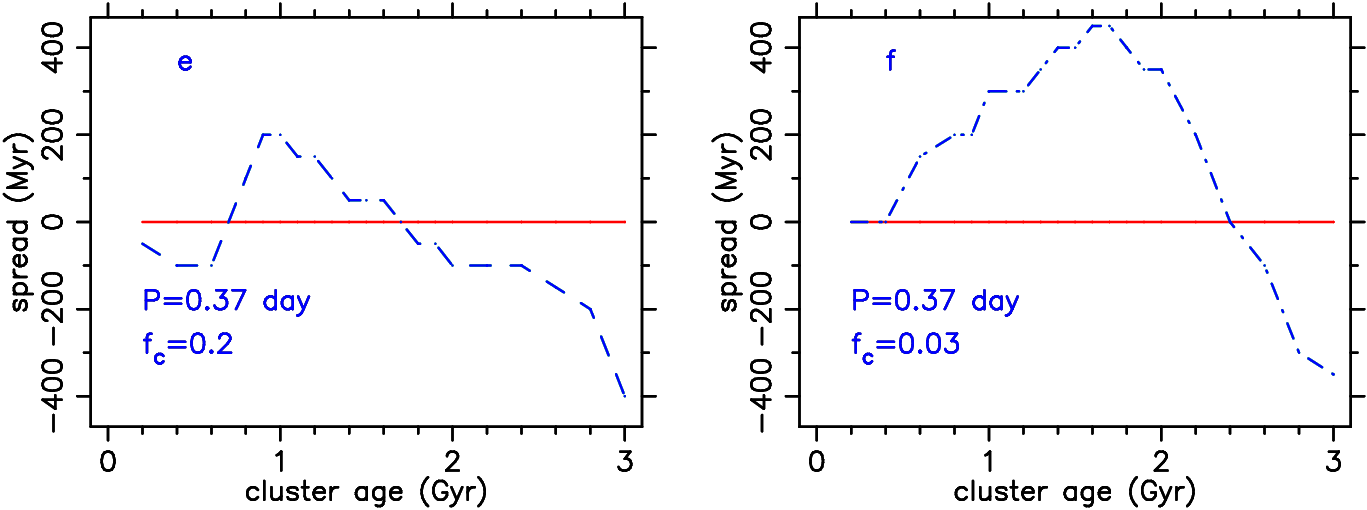}
\caption{The dashed lines show the equivalent spread of the MSTO of
star clusters caused by rotation. The positive and negative spreads
represent the fact that the MSTO of rotating models are redder and bluer
than that of non-rotating ones with the same age respectively.
 \label{fsp}}
\end{figure}

\begin{figure}
\includegraphics[angle=-90, scale=0.35]{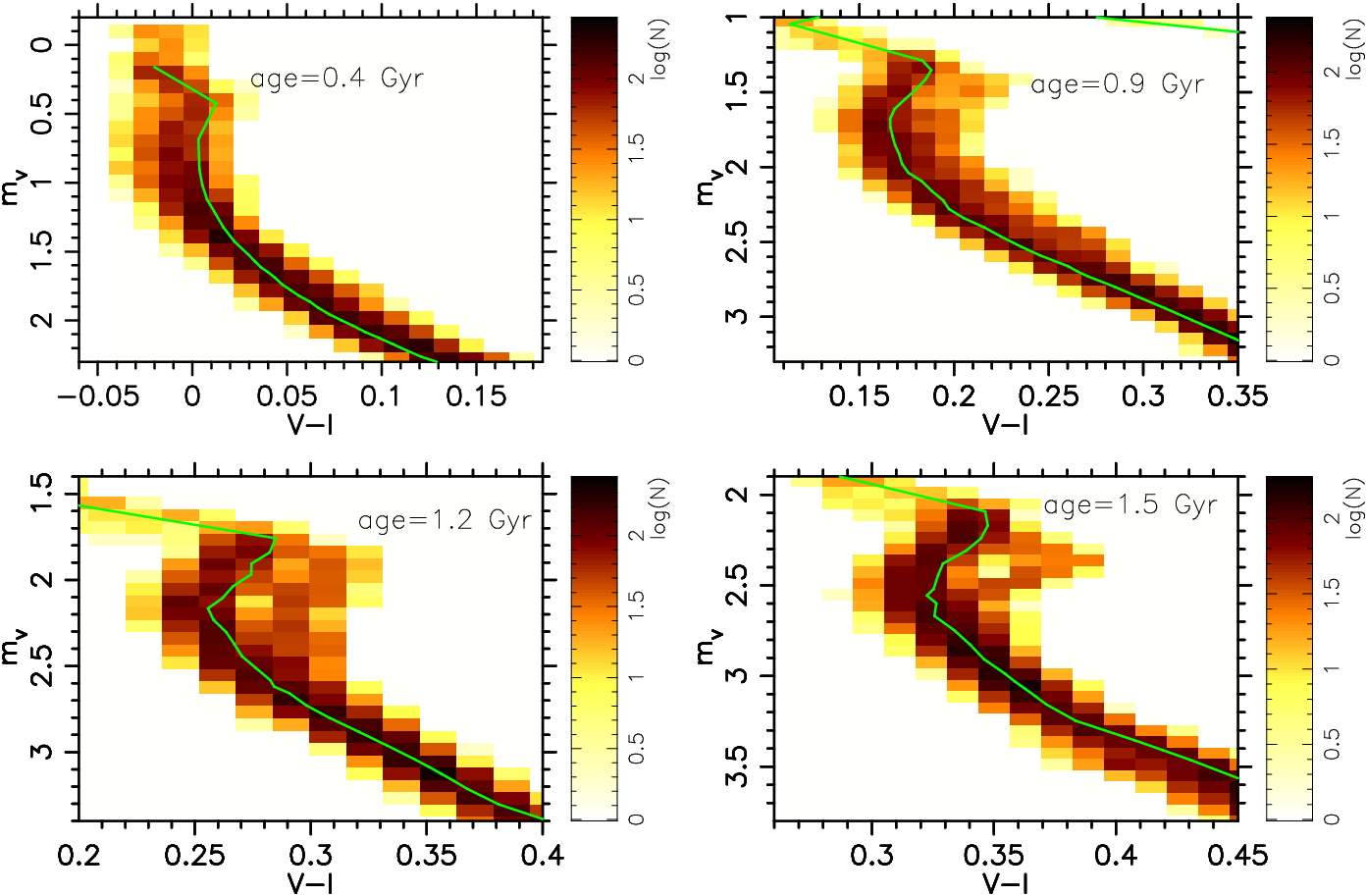}
\includegraphics[angle=-90, scale=0.35]{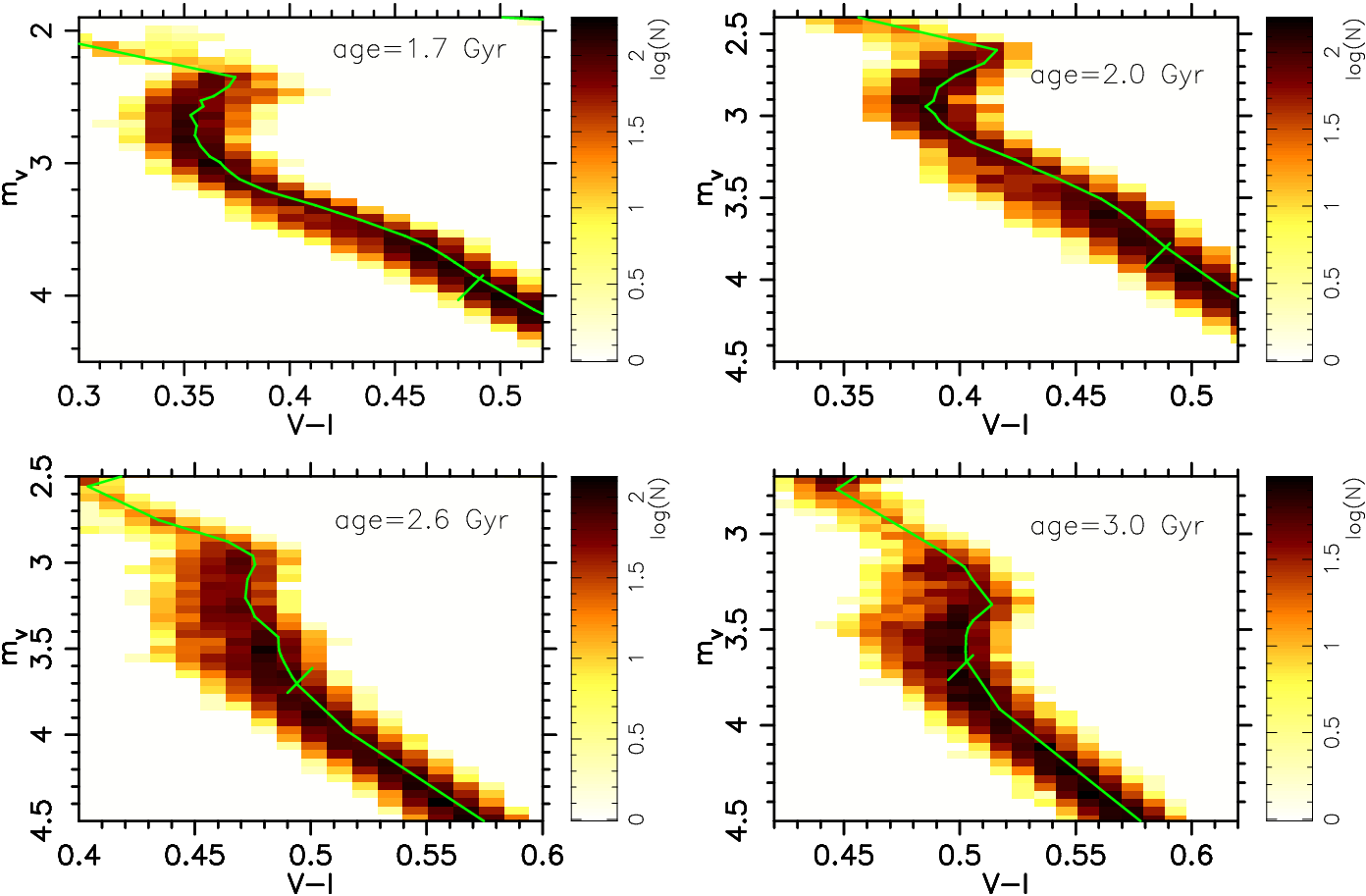}
\caption{The CMDs of the synthetic star clusters at different ages.
The total mass is assumed to be of about $3\times 10^{4}$ \dsm{}, following
a \cite{chab01} log-normal initial mass function. We assumed 30\% of stars
rotating with $P_{0} = 0.49$ day and $f_{c}= 0.2$. The
solid (green) line shows the isochrone of non-rotating models.
The position of the model with $M=1.15$ \dsm{} is marked by a slash.
The typical errors in magnitude and color are 0.01 and 0.015 respectively.
 \label{fvg}}
\end{figure}

\begin{figure}
\includegraphics[angle=-90, scale=0.34]{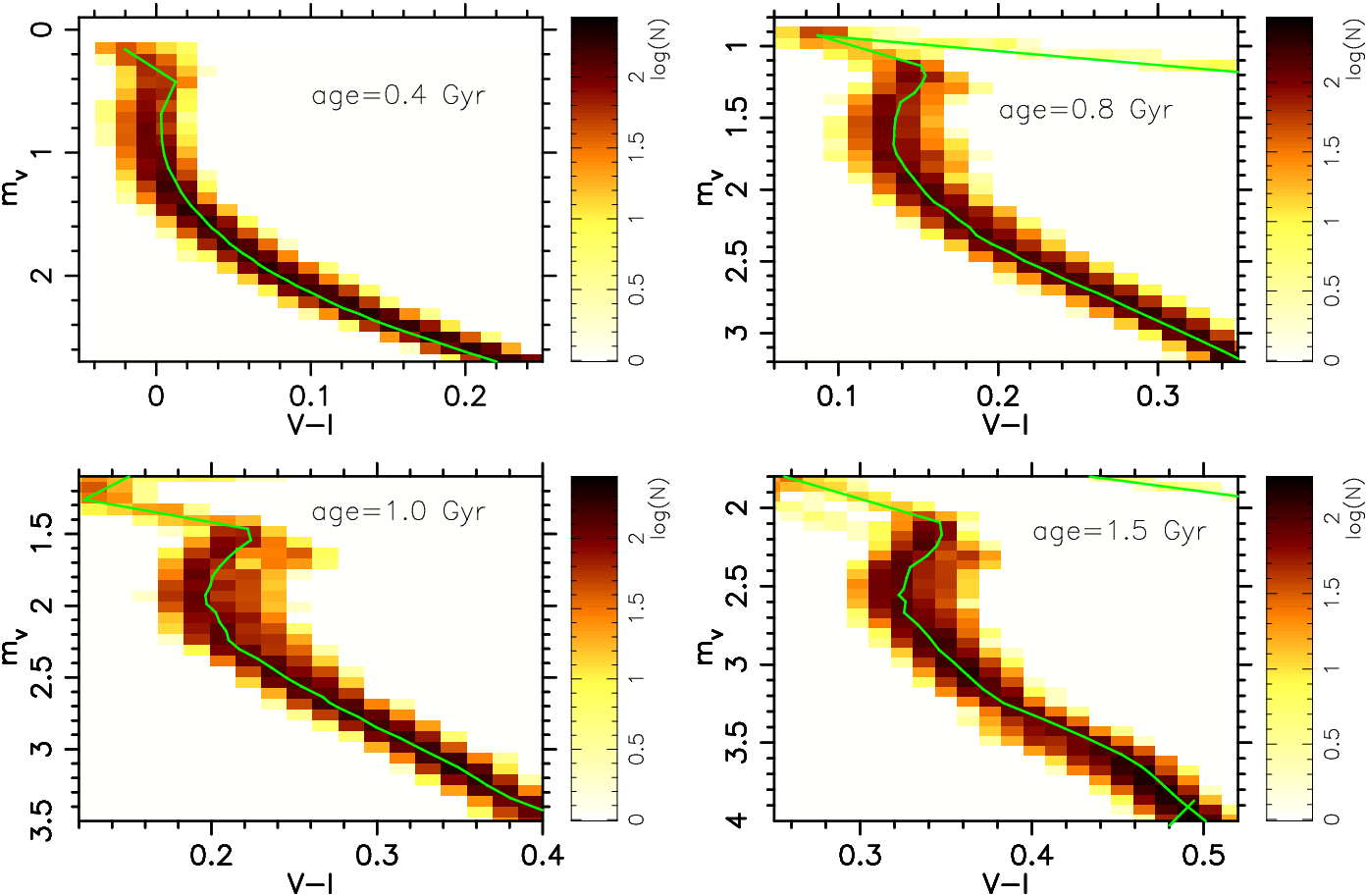}
\includegraphics[angle=-90, scale=0.34]{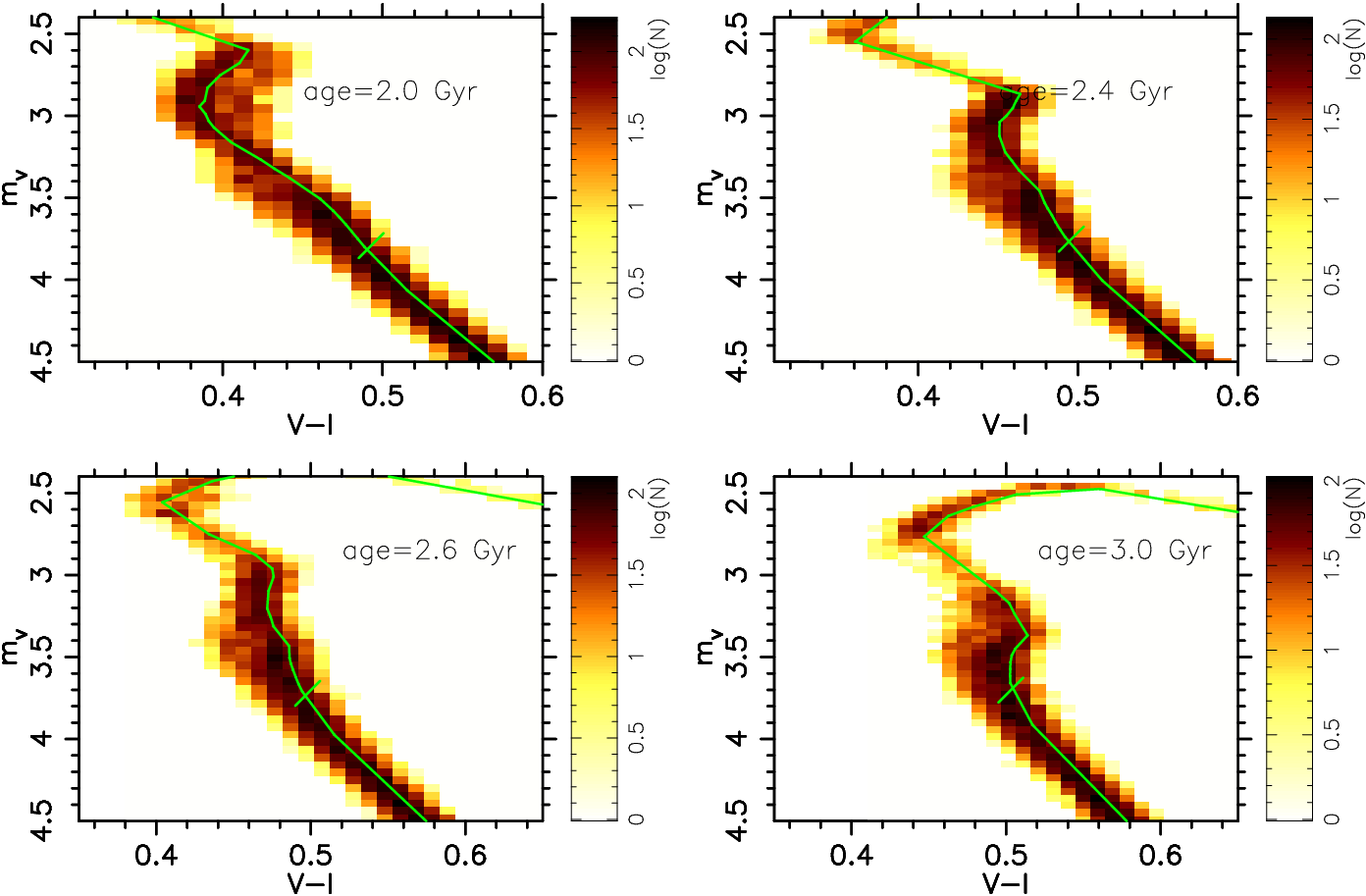}
\caption{Same as Fig. \ref{fvg}, but the initial period of rotating
population here is 0.73 day. \label{fvgl}}
\end{figure}

\begin{figure}
\includegraphics[angle=-90, scale=0.35]{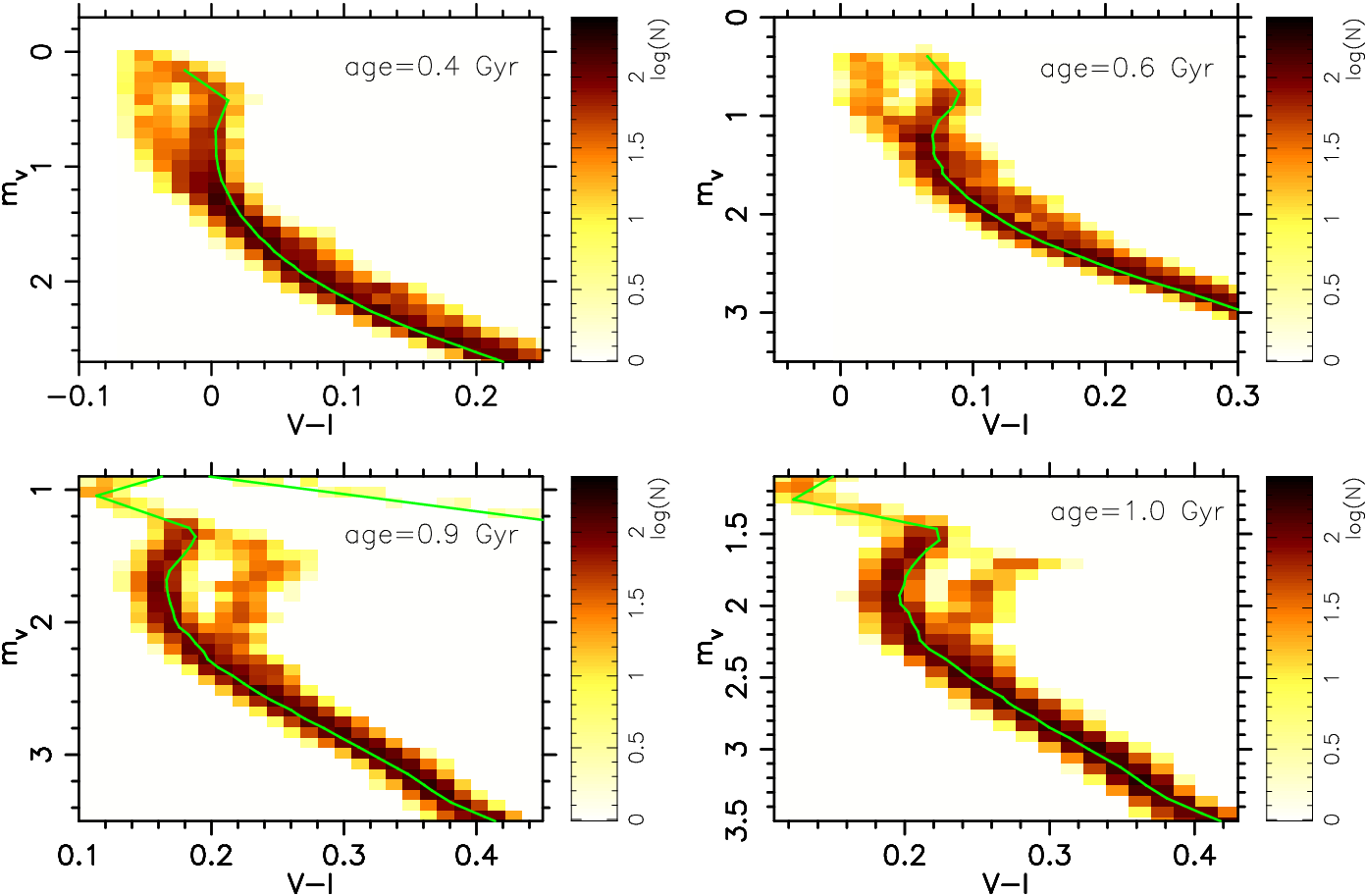}
\includegraphics[angle=-90, scale=0.35]{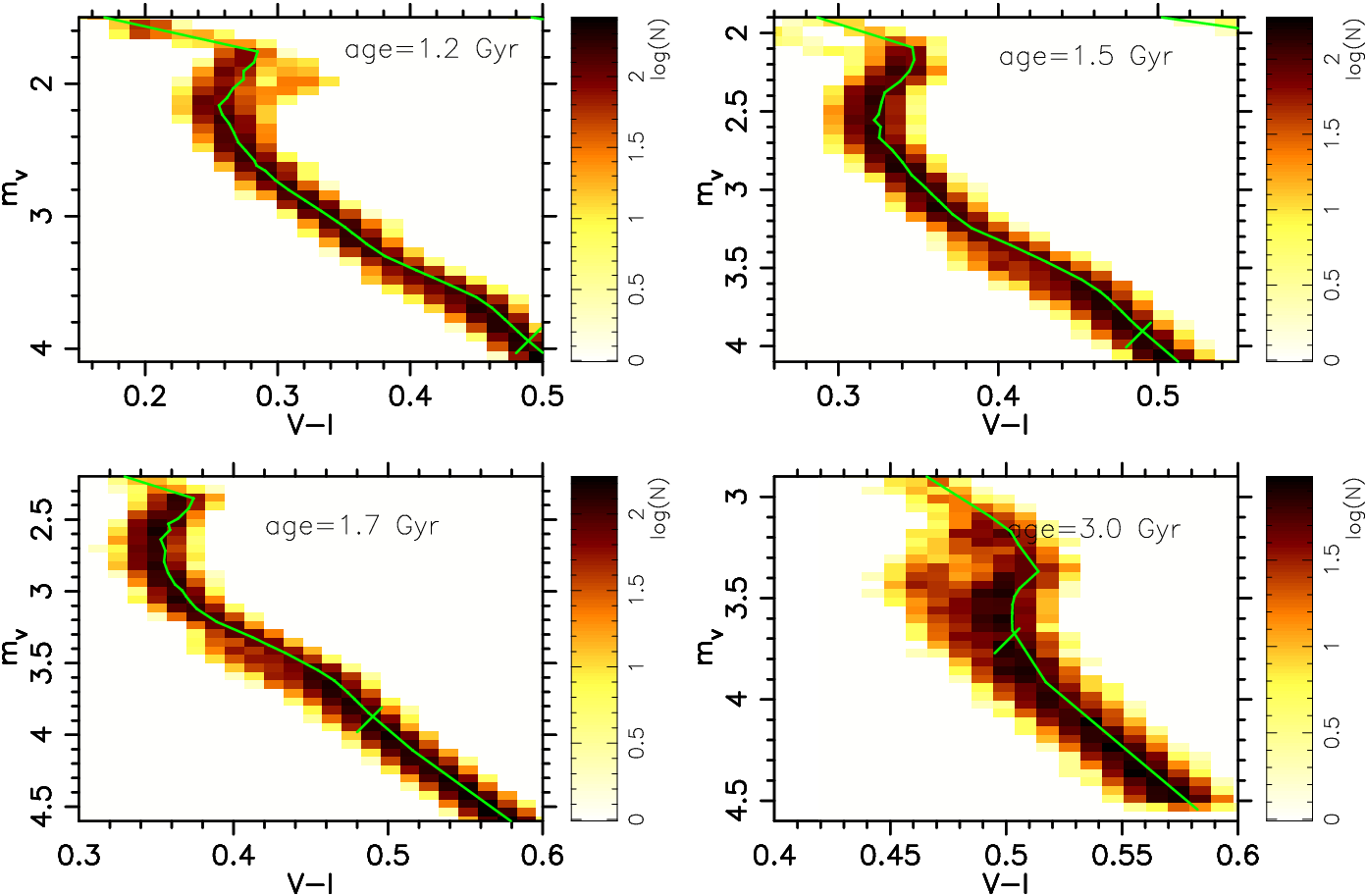}
\caption{Same as Fig. \ref{fvg}, but the initial period of rotating
population here is 0.37 day. \label{fvgh}}
\end{figure}

\begin{figure}
\includegraphics[angle=-90, scale=0.35]{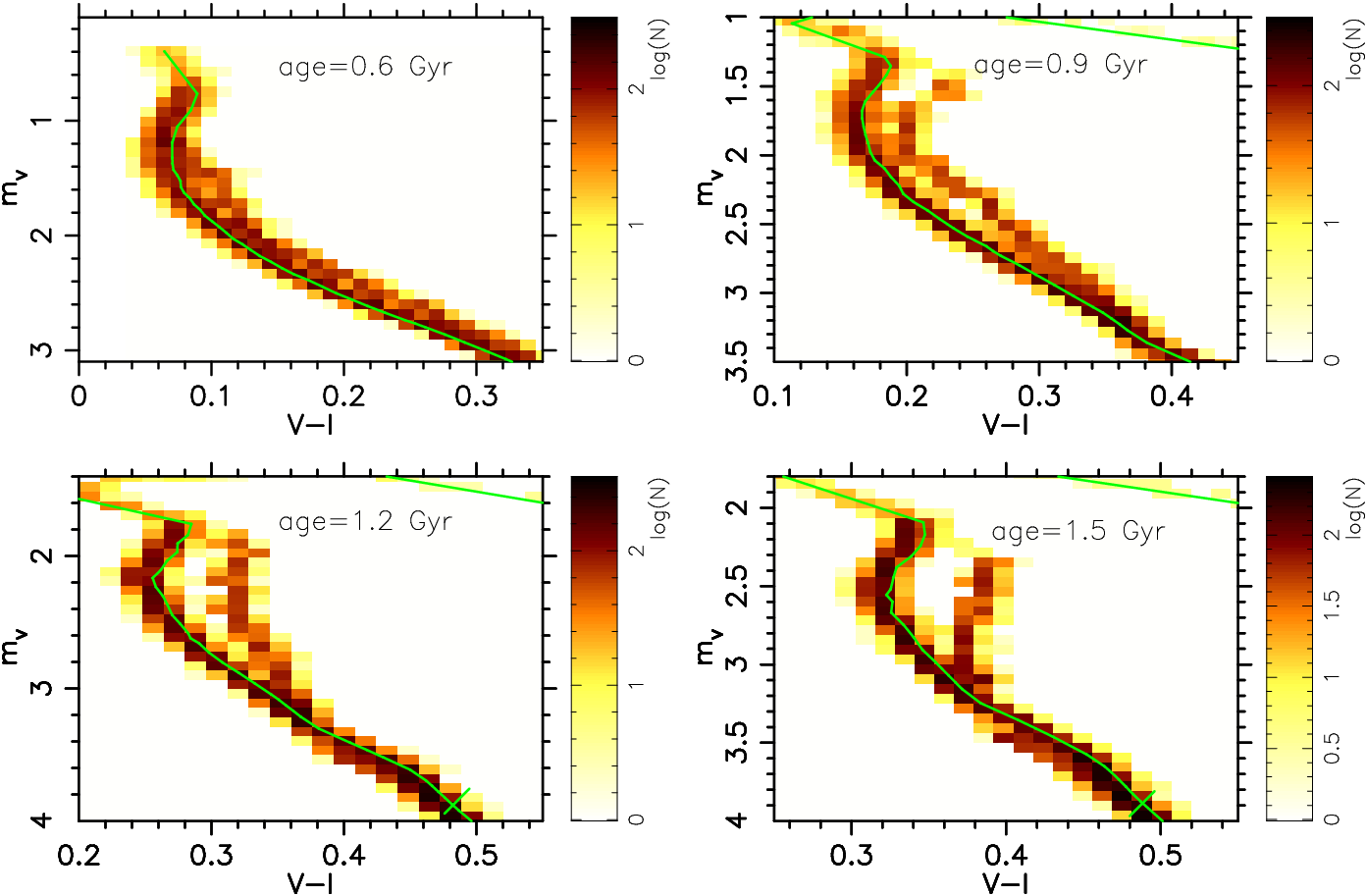}
\includegraphics[angle=-90, scale=0.35]{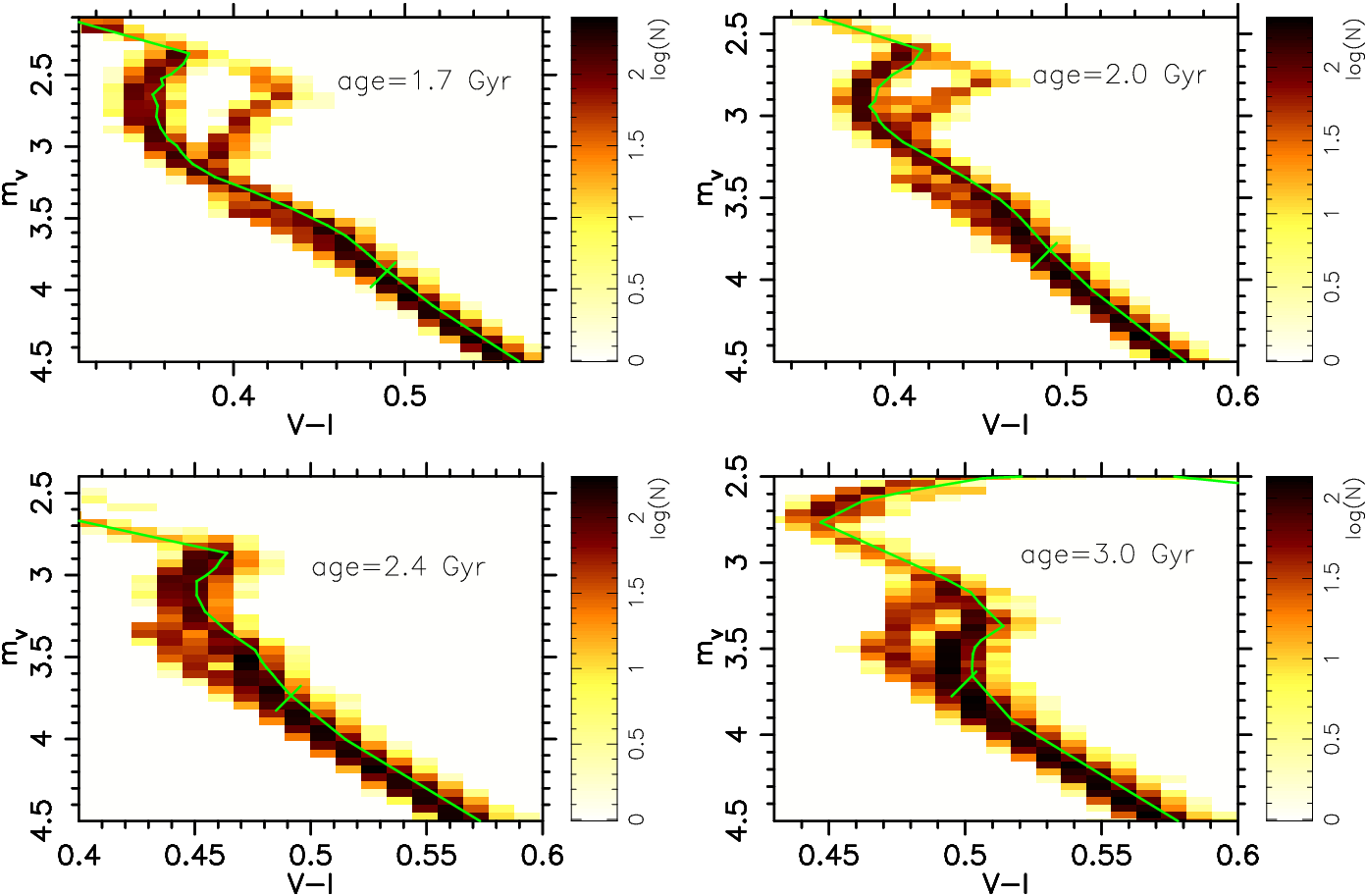}
\caption{Same as Fig. \ref{fvg}, but the value of $f_{c}$
is 0.03. \label{fvgf}}
\end{figure}

\begin{figure}
\includegraphics[angle=-90, scale=0.5]{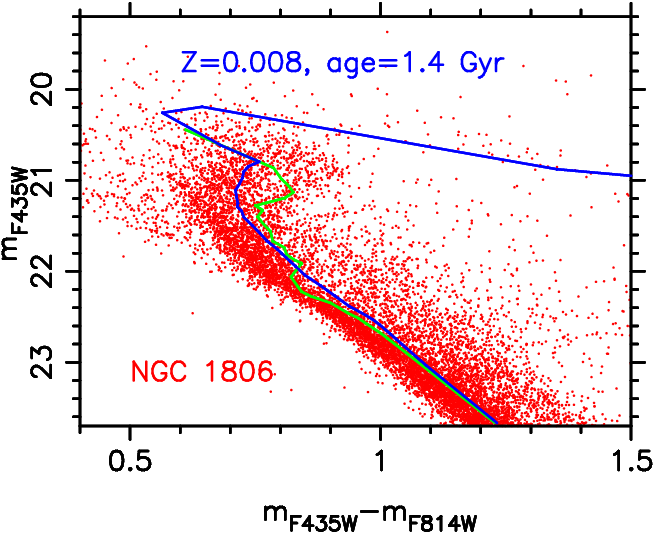}
\includegraphics[angle=-90, scale=0.5]{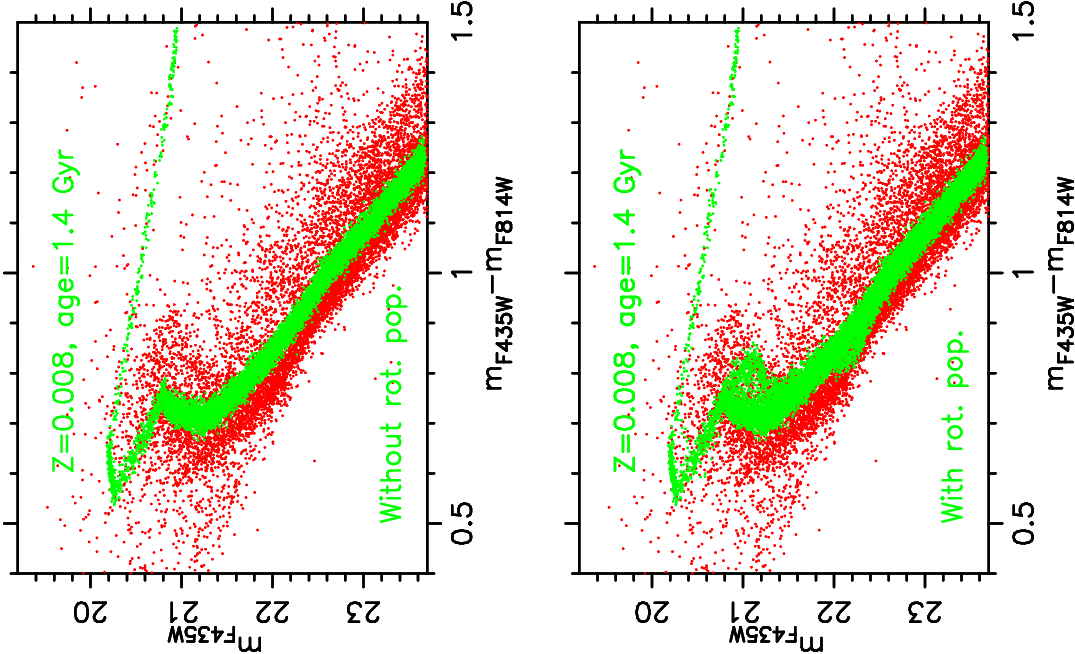}
\caption{The CMDs of the star cluster NGC 1806. The red
points indicate the observed data \citep{milo09}. The
green points show the simulated data. A distance modulus of 18.45
is adopted. The value of 0.14 is adopted for
$E(m_{\mathrm{F435}}-m_{\mathrm{F814}})$, which is larger than the 0.09
given by \cite{milo09}. The blue and green lines in the top panel
show the isochrones of non-rotating and rotating models with
an age of 1.4 Gyr. The jump of the green line at $m_{\mathrm{F435}}\approx22.1$
is not significant in (V-I) (see Fig. \ref{fcmd}) and is related to the
absence of magnetic braking in stars with $M>1.3$ \dsm{}.
The simulated population shown in the middle
panel does not include rotating models, but those shown in
the bottom panel include a 30\% rotating population obtained
by interpolating between rotating models with different
rotation rates.
\label{fngc}}
\end{figure}

In this work, we calculated a grid of evolutionary tracks of
rotating and non-rotating stars with $Z=$ 0.008 and
masses between 1.0 and 3.0 \dsm{}. For a given rotation rate,
the effects of rotation on the structure and evolution of the
stars are dependent on their mass and age. For
the initial rotation period of 0.49 day and the efficiency of
mixing of 0.2, rotating models have a fainter and redder turnoff
with respect to their non-rotating counterparts when age is
located between about 0.8 and 1.9 Gyr. When the age is larger than
2.4 Gyr, rotating models have a brighter and bluer turnoff
with respect to the non-rotating ones. However, when the
age is less than 0.6 Gyr, the rotation hardly affects
the MSTO of the star clusters. The extension of the MSTO caused
by the rotation can be equivalent to an age spread of about
200 Myr when the age of the star clusters is located between
about 1.2 and 1.6 Gyr. The extension decreases with the decrease
or increase in age \textbf{(see the panel $c$ of Fig. \ref{fsp})}.
When the initial rotation period decreases to about 0.37 day,
the extension of the MSTO caused by the rotation disappears
in star clusters with ages between about 1.4 and 2.0 Gyr
\textbf{(see the panel $e$ of Fig. \ref{fsp})}.
Young star clusters with age less than 0.7 Gyr
have a blueward extended MSTO; but star clusters with
ages between 0.8 and 1.3 Gyr still have a redward
extended MSTO. When the initial rotation period increases
to about 0.73 day, the MSTO of star clusters with age less than
0.7 Gyr cannot be affected by rotation, but the MSTO of
star clusters with ages between 0.9 and 2.2 Gyr extends
redward \textbf{(see the panel $a$ of Fig. \ref{fsp})}.
When the value of the efficiency of mixing decreases
to 0.03, the extension of the MSTO caused by rotation can be
equivalent to the effect of an age spread of about 400 Myr
\textbf{for intermediate-age star clusters (see the panels $d$
and $f$ of Fig. \ref{fsp})}. On the whole,
the rotationally induced extension of the MSTO of intermediate-age
star clusters is dependent on the rotation rate, the age of star clusters,
and the efficiency of the mixing of elements. But the blueward
extension of the MSTO of star clusters with age larger than
about 2.4 Gyr is mainly dependent on the efficiency of the mixing.

Our simulation shows that rotation could lead to a double or
extended MSTO in star clusters with ages between 0.8 and 2.2 Gyr or
larger than 2.4 Gyr. The rotation seems not to be able to result in
the extension of the MSTO of star clusters with age less than 0.6 Gyr.
At least, the extension of the MSTO of these star clusters is not
significant, as compared to that of intermediate-age star clusters.
Considering the difference in the initial rotation rates and the efficiency
of mixing, the redward extended MSTO may be able to be more easily observed in
star clusters with ages between about 0.9 and 1.5 Gyr.


\acknowledgments We thank the anonymous referee for helpful comments,
A. P. Milone for providing the data of NGC 1806, and Daniel Kister
for help us in improving English; and we acknowledge the support from the
NSFC 11273012, 11273007, 10933002, and 11003003, and the Project
of Science and Technology from the Ministry of Education (211102).

\clearpage

\clearpage

\clearpage

\end{document}